\documentclass{iaus}
\usepackage{graphicx}

\title[Global 3D MHD Simulations of Accretion Disks]%
{New Global 3D MHD Simulations of Black Hole Disk Accretion and Outflows}

\author[Dobbie et al.]  %
{Peter B. Dobbie$^1$, Zdenka~Kuncic$^1$, \break Geoffrey~V.~Bicknell$^2$ and Raquel~Salmeron$^2$}

\affiliation{$^1$School of Physics, University of Sydney, NSW 2006, Australia \break email: p.dobbie@physics.usyd.edu.au\\[\affilskip]
$^2$Research Centre for Astronomy and Astrophysics, Australian National University, ACT 2611, Australia}

\pubyear{2009}
\volume{259} %
\pagerange{100--100}
\date{"9 December 2008" and in revised form ??}
 \jname{Cosmic Magnetic Fields: From Planets, to Stars and Galaxies} \editors{K.G. Strassmeier, A.G. Kosovichev \& E. Beckman, eds.}

\begin{document}

\maketitle

\begin{abstract}
It is widely accepted that quasars and other active galactic nuclei (AGN) are powered by accretion of matter onto a central supermassive black hole. While numerical simulations have demonstrated the importance of magnetic fields in generating the turbulence believed necessary for accretion, so far they have not produced the high mass accretion rates required to explain the most powerful sources. We describe new global 3D simulations we are developing to assess the importance of radiation and non-ideal MHD in generating magnetized outflows that can enhance the overall rates of angular momentum transport and mass accretion.
\keywords{accretion, accretion disks --- magnetic fields --- (magnetohydrodynamics:) MHD --- radiative transfer --- turbulence --- methods: numerical --- galaxies: active, jets, magnetic fields, nuclei --- (galaxies:) quasars: general --- X-rays: binaries}
\end{abstract}

\section{\label{sec:1}Introduction}
Accretion onto a black hole is widely believed to power the central engine of high-energy astrophysical objects such as active galactic nuclei (AGN) and some X-ray binaries. Since the standard theory of astrophysical disk accretion was formulated over 30 years ago (\cite{sha73,novthorn73}), arguably the most important advances in understanding these systems have come from computational modelling.  Numerical simulations demonstrate unequivocally that the magnetorotational instability (MRI) can produce magnetohydrodynamic (MHD) turbulence and the outward transport of angular momentum required for accretion to proceed (see \cite{bal03} for a review).

Notwithstanding the important advances made using MHD simulations over the last two decades, they have so far been unable to resolve two key outstanding issues:
1. How are the high mass accretion rates inferred in the most powerful sources achieved? and
2. How are the outflows and jets observed across the mass spectrum of accreting sources produced?
Our new work will test the hypothesis that these issues may be connected and mutually resolved by a generalized model for MHD disk accretion.

\section{\label{sec:2}MHD Disk Accretion Theory}
\begin{figure*}[t]
\centering
  \includegraphics[width=0.9\textwidth]{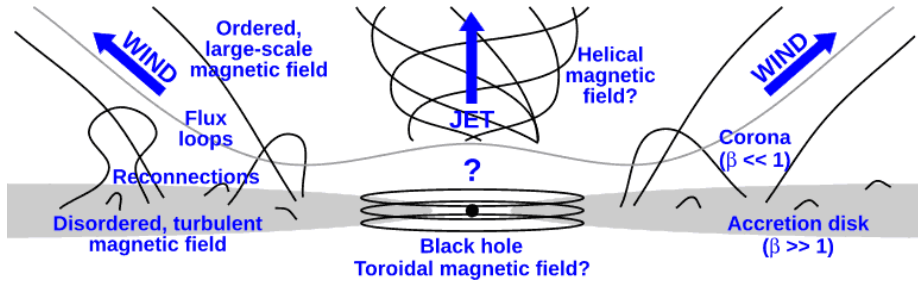}
  \caption{A schematic illustration of the inner regions of an MHD accretion flow around a black hole. $\beta$ is the usual plasma beta (the ratio of gas to magnetic pressures). The wind can drive mass loss from the disk, while the jet may be dominated by Poynting flux.}
  \label{fig:bhaccretion}
\end{figure*}

The analytical foundation for our new work is based on the generalized accretion disk model of \cite{kun04} which is schematically illustrated in Fig.~\ref{fig:bhaccretion}. The model includes contributions from both radial and vertical transport of angular momentum to the overall mass accretion rate: angular momentum is transported radially outwards by internal MHD stresses and vertically outwards by both mass outflows in a wind and MHD stresses acting over the disk surface.  This vertical transport of energy and momentum is not modelled in standard accretion disk theory.  Our new simulations are designed to compare its effects with the radial turbulent transport of angular momentum and to test the prediction of a modified emission spectrum (see e.g. \cite{kunbick07ab}).

\section{\label{sec:3}New MHD Accretion Simulations}

To implement our model, we are using and extending FLASH\footnote{FLASH is freely available at http://flash.uchicago.edu.} (\cite{fry00}), the public MHD code developed at the University of Chicago.  FLASH implements adaptive mesh refinement and solves the equations of MHD in conservative form thus explicitly conserving energy.

To date, most simulations of the MRI in accretion disks have been conducted in a shearing box approximation but this approach introduced complications and pitfalls (see e.g. \cite{reg08}).  We will instead conduct global 3D simulations.

By explicitly modelling a finite resistivity, we will explore its effect on the evolution of the magnetic field topology, particularly the self-consistent emergence of a significant $z$-component necessary to produce outflows and high mass accretion rates.

We will include radiation in our simulations to directly compare our results to the observational data. This will allow us to test whether the blackbody emission from the disk is modified by outflows as predicted. Radiative transfer is also be required to transport the internal energy dissipated in the disk plasma at the end of a turbulent cascade, i.e. to cool the disk.  In addition, it may be that regions of the disk where radiation dominates may be thermally unstable (\cite{sha76}), thus affecting the dynamics.

\end{document}